\documentclass[12pt]{article}

\usepackage{amsmath,amssymb,cite}

\newcommand{\eqna}[1]{\begin{subequations} \label{#1}
\begin{eqnarray}}
\def\eena{\end{eqnarray}
\end{subequations}}

\def\beq{\begin{equation}}
\def\eqn#1{\beq\label{#1}}
\def\ee{\end{equation}}

\def\bb {\begin {eqnarray}}
\def\eqnn#1{\bb\label{#1}}
\def\eea {\end {eqnarray}}

\def\nn{\nonumber}
\input amssym.def
\input amssym.tex

\def\({\big(}
\def\){\big)}

\def\a{\alpha}
\def\b{\beta}
\def\g{\gamma}
\def\d{\delta}
\def\e{\epsilon}

\def\l{\lambda}

\def\r{\rho}

\def\s{\sigma}

\def\G{\Gamma}

\def\O{\Omega}

\def\o{\omega }

\def\tri{\triangle}

\def\zu{\Upsilon_b}

\def\la{\langle} \def\ra{\rangle}
\def\zL{\Lambda}
\def\IR{{ \Bbb R} }
\def\IZ{{ \Bbb Z} }
\def\IC{{ \Bbb C} }

       \def\CZ{{\cal Z}}

 \def\bh{\bar h}
\begin{document}

\textheight=24cm
\textwidth=16.5cm
\topmargin=-1.5cm
\oddsidemargin=-0.25cm.

\input amssym.def
\input amssym.tex
\def\IZ{\Bbb Z}\def\IR{\Bbb R}\def\IC{\Bbb C}\def\IN{\Bbb N}
\def\gg{\goth g} \def\bgo{\overline{\goth g}}
 \def\bgh{\bar{\goth h}}
 \def\gp{\goth p}\def\hp{\overline{\goth p}}
\def\gh{\goth h}
\def\ga{\goth a}
\def\gk{\goth k}
\def\gb{\goth b}

\def\zC{\Gamma} \def\zD{\Delta} \def\zF{\Phi} \def\zL{\Lambda}
 \def\zO{\Omega} \def\zT{\Theta}

 \def\bdot{\star}
\def\tb{\tilde{\beta}}
\begin{center}

{\LARGE {\bf $W_4$ Toda example as  hidden Liouville CFT
}\footnote{Invited talk given by  V.B. Petkova at the
International Workshop "Supersymmetries and Quantum Symmetries", 
Dubna,  August 3-8, 2015.}
}

 \vspace{10mm}
 
 {\bf \large  P. Furlan$^{*}$} and  {\bf  \large V.B. Petkova$^{**}$} 
 \vskip 5mm

\emph{
$^{*)}$Dipartimento  di Fisica 
 dell'Universit\`{a} di Trieste, Italy\\
 furlan@ts.infn.it\\
$^{**)}$Institute for Nuclear
Research and Nuclear Energy, \\
 Bulgarian Academy of Sciences,
% 72   Tsarigradsko Chaussee, 1784 
 Sofia, Bulgaria\\
petkova@inrne.bas.bg
}

\end{center}

\vskip 4mm

\begin{abstract}

We construct  correlators in the $W_4$ Toda 2d conformal field theory  for a particular class of
representations  and demonstrate a   
relation to a 
$W_2$ (Virasoro)  theory with different central charge. The relevance of the classical  limits of the constructed  3-point functions and braiding matrices  to problems in 4d  conformal theories 
is discussed.

\end{abstract}

\section{Introduction}
\setcounter{equation}{0}
There are few explicit results for the basic structures of the  2d conformal field theories  
(CFT) based on  higher rank algebras, such as the WZW models,  or their  quantum Drinfeld-Sokolov reductions,  the  Toda CFT. The main motivation of this work is to extend the examples in the literature 
\cite{FLc, FLd} (see also \cite{Wyl, IMP, BEFS}),  mostly for the $W_3$ case, 
for a  particular $W_4$ Toda CFT. The latter is   selected by   the  possible applications to the 4d models with superconformal symmetry $sl(2,2|4)$ in the context of the AdS$_5$/CFT$_4$ correspondence.

The talk is based mostly on \cite{FP} . 
In Section 2 we exploit  the Coulomb gas construction technique of \cite{FLc,FLd} to compute
the 3-point functions for  a class of  representations of the $W_4$ algebra and then we discuss their 
light charge classical limit comparing  with computations in the supergravity approximation to the string theory.
%  in the  AdS$_5$/CFT$_4$ correspondence.
 The 3-point constants  are   used in Section 3  to solve the crossing symmetry equation
for  the braiding matrices of the 4-point conformal blocks in which one of the fields is labelled by the 
highest weight of the 6-dimensional fundamental $sl(4)$ representation. 
In Section 4  the  local 4-point function is shown   to admit an  explicit  integral representation. Surprisingly,  the latter  is identified with a  4-point Liouville  correlator 
with one degenerate field. The vertex operators are described by representations of the Virasoro algebra  with modified central charge,  which differs from  
the  central charge  of the Virasoro subalgebra of the $W_4$ algebra. This relation,  announced in \cite{FP}, 
is demonstrated here for  the fusing matrices of the two theories. In the last Section 5 we discuss the relevance  of the  heavy charge classical limit  for a braiding identity and its solutions  for the study of the quasiclassics of sigma  models in the context of the AdS/CFT correspondence.

\section{3-point functions}
\setcounter{equation}{0}

We consider the $W_4$ CFT with central charge parameterized by  real $b$
\eqn{tod}
c_T= 3 (1+20 Q^2)=3\(41+ 20(b^2+{1\over b^2})\)\,,   
\ Q={1\over b}+b\,.
\ee
The scalar primary fields  $V_\b(z, \bar z)$  are  labelled by a generic $sl(4)$ weight in the free field (Coulomb gas)
 representation, see  \cite{FLuk, FLc} for details. Equivalent representations carry  charges related by the action of the Weyl 
reflection group
\eqn{weyltod}
w\bdot \b= 
Q \r +w(\b-Q\r )\, 
\ee
and the operators are related by a reflection amplitude 
$$
V_{ \b}(z,\bar z) = R_w(\b) V_{w^{-1}\bdot \b}(z,\bar z)
$$
The action \eqref{weyltod} preserves the three quantum numbers 
characterising the $W_4$ representations,  in particular,  the 
conformal dimension 
 \eqn{dimqr} 
\tri(\b)= {1\over 2}(\b, 2\r Q-\b)\,.   
 \ee
Here $ \ \r =\sum_{i=1}^3 \o_i$ is the  Weyl vector and $\o_i\,, i=1,2,3$ are the three $sl(4)$  findamental weights
$(\o_i, \a_j)=\d_{ij}$ for $\a_j$ denoting the $sl(4)$ simple roots.

 We start with computing the 3-point OPE constant  for a particular set of  symmetric $\b_a=\b_a^*$  highest weights
\eqna{classreps}
&&(\b_a, \a_1)=0= (\b_a, \a_3)\,, \ {\rm for} \ a=1,2  \,, \\
&& \ (\b_3,\a_1)=(\b_3,\a_3)\,.
 \eena
The  weights  (\ref{classreps}a) correspond to scalars in 
 the context of  the 4d  conformal group  
  representations, with 4d conformal dimension $\tri=(\b, \o_2)/b  $.

To compute the 3-point constant  we exploit the technique of  \cite{FLc} based on the Baseilhac - Fateev  (BF) integral relation 
\cite{BF}. It  allows to derive  a 
 recurrence relation for the  
corresponding Coulomb gas integrals in which the three charges are restricted by a charge neutrality condition.
The solution is then analytically continued similarly to
the Liouville case with the result
  \eqnn{formT} 
&& C(\b_1, \b_2, \b_3)  \nn \\
&&= \(b^{2(1-b^2)}\pi \mu \g(b^2)\)^{ {(2\r Q- \b_{123},\r)\over b}}\,  
{\prod_{\a=\a_1,\a_{14}}\zu((\b_3-\r Q,\a)+Q)\over \prod_{\a=\a_1,\a_{13}}\zu((\b_3-\r Q,\a))}\times \\
&&\prod_{a=1,2}{\zu((\b_a,\a_2))
\over  \zu((\b_{123}-2\b_a,\o_2-\o_1)) }{\zu(b)\, \zu((\b_3,\a_2))   \over  \zu((\b_{12}^3,\o_1)) \zu((\b_{123},\o_2-\o_1) -2Q) }\times  
\cr
&&\prod_{a=1,2}{  \zu((\b_a-\r Q,\a_{24})+Q)  \over   
\zu((\b_{123} -2\b_a,\o_1)-Q)} {\zu(b)\, \zu((\b_3-\r Q,\a_{24})+Q) \over  
 \zu((\b_{12}^3,\o_2-\o_1)-Q) \zu((\b_{123},\o_1)-3Q) }\,.\nn
\eea
 Here $\b_{123}=\sum_{a=1}^3 \b_a\,, \b_{12}^3=\b_{123}-2\b_3$.
Recall that 
 $\zu(x)$ is an entire function with zeros at $x=-nb-m/b\, $ and $x=Q+nb+m/b$, $ n,m\in \CZ_{\ge 0}\,, $ satisfying  
the functional  relations
\eqn{fnct}
\zu(x+b^{\epsilon})=  \g(xb^{\epsilon})\,b^{\e(1-2x b^{\e})} \,\zu(x)\,, \epsilon=\pm 1\,, \ \g(x)=\G(x)/ \G(1-x)\,.
\ee
 We introduce notation for the  Coulomb gas OPE constant  reproduced as a double  residue
\eqnn{resdoubl} 
&&c(\b_1, \b_2, 2\r Q- \b_3):= \\
&& {\rm res}_{ (\b_{12}^3, \o_2-2\o_1)/b=-(s_2-2s_1) } \, {\rm res}_{(\b_{12}^3,\o_1)/b=-s_1}C(\b_1, \b_2, 2\r Q- \b_3)
\nn
\eea
where $s_1$ and $s_2-2s_1$ are nonnegative integers. In particular if $\b_1=-b \o_2$,  corresponding to the
highest weight $\l=\o_2$ of the 6-dim  $sl(4)$ representation, the formula  \eqref{resdoubl} reproduces the 
structure  constants  
 of the fusion of the  fundamental field  $ V_{-\o_2 b}$  and  $V_{ \b}$  
  corresponding to  the shifts  with three  of the six  points of  the weight diagram $\Gamma_{\o_2}$ 
\eqnn{weidi}
&&c_h(\b):=c(-\o_2 b, \b, 2\r Q- (\b- h b))=\la V_{-\o_2 b}(0)V_{ \b}(1) V_{2\r Q- (\b-h b)}\ra \,, \nn \\
&& h=\pm \o_2\,, \ 
 \o_1+\o_3-\o_2
=: \bh\,. 
\eea
 In particular $c_{h=\o_2}(\b)=1$.
 The remaining three OPE constants, computed  in general in \cite{FLc}  vanish,  in agreement with the vanishing of the inherited from the 
 $sl(4)$  tensor product decomposition  multiplicities for the particular partially degenerate weight 
    (\ref{classreps}a).

Consider the special case when all three weights in \eqref{formT} are of the type (\ref{classreps}a).
 It is instructive to display the asymptotics of \eqref{formT}  in the classical limit $b\to 0$ with   "light"  charges, i.e., $(\b_a, \a_2)/b=\sigma_a $   remain   finite. More precisely, consider the modified 3-point constant 
 $\bar{C}(\b_1, \b_2, \b_3 )$
  in which all   $Q$- factors are replaced by $Q\to b$. This is achieved by factorizing 
a  finite number of   $\g(z+{1\over b})$ - factors extracted by applying the functional relation \eqref{fnct}. 
Then the   light charge classical limit reads
\eqnn{cllm}
&&\bar{ C}(\s_1 b \o_2, \s_2 b \o_2, \s_3 b \o_2)
   \sim  \\
&&\G({\s_{123}\over 2}- 2) 
\prod_{a}{\G({\s_{123} \over 2} - \s_a) \over \G(\s_a)   }\,
 \G({\s_{123}\over 2}- 3) 
\prod_{a}{\G({\s_{123} \over 2} - \s_a-1) \over \G(\s_a-1)   }\,. \nn
\eea
This expression 
 reproduces, up to field normalisation,  the 4d scalar 3-point correlators  defined
as an integral over 
AdS$_5$ of three  bulk-boundary kernels representing the   vertex operators 
in the supergravity approximation to string theory \cite{FMMR}. 
This result suggests that  the modified 3-point constant  $\bar{C}(\b_1, \b_2, \b_3 )$ describes the WZW  counterpart of the Toda 3-point  constant;  see  \cite{FGPP} for a discussion
of the quantum 
Drinfeld - Sokolov reduction on the level of correlators. 

 Analogous to \eqref {cllm} classical limit arises from the "matter" counterpart  in the dual central charge region with the parameter $b \to i b$. It reproduces the integral over $S^5$,  see \cite{FP} for more details  and a further discussion of a  BPS type relation for the charges in the two dual regions,  leading to a trivial 3-point constant of the corresponding  4d superconformal  correlator.

\section{ 4-point function and  fusing  matrix} 
\setcounter{equation}{0}

Our next step is to consider the local 4-point function $\la V_{f} V_{\b_1}V_{\b_2} V_{\b_3} \ra\, $ of  primary spinless operators $V_{\b}(z,\bar z)$ one of which is labelled by  a fundamental highest weight, 
in our case $f=-b\o_2$. All the other three weights $\b_a\,, a=1,2,3$ are supposed   to be in  the class (\ref{classreps}a)  labelling  doubly reducible Verma modules with at most two singular vectors  since $(\b_a, \o_2)$ 
are assumed to take generic values.

The standard approach to the study of the 4-point functions with degenerate fields is to exploit the differential equations for  the chiral blocks arising from the factorization of all singular vectors (and possibly some descendent states) along with the restrictions imposed by the Ward identities of the $W_4$ algebra. In higher rank  CFT this is a difficult technical task,  see \cite{BW} for an early discussion of the peculiarities of the Toda correlators. 
On the other hand we expect that 
in the particular  example of several 
reducible representations 
the set of restrictions will be sufficient  to determine 
 the chiral blocks and furthermore to reduce the 
 fusion channels  to the ones 
  determined 
 by the OPE coefficients of the primary fields with the fully degenerate field   $V_{-\o_2 b}$ \eqref{weidi}. 
 In the next section we shall confirm this expectation by deriving an explicit alternative integral representation for  the 4-point function.

 $\bullet$ 
The 4-point function admits different  equivalent  diagonal  decompositions in   conformal blocks. 
They are    related by  linear  transformations, realizing the action of the elements of 
braiding group with generators
$e_i\,, i=1,2,3$ on the plane (Riemann sphere)  with 4 holes;  
$e_i $ is exchanging the chiral vertex  
 operators at the $i$-th and $i+1$-th points.
 In particular the generators  $e_2$
 (for the above order of the corresponding chiral vertex operators) is represented by non-trivial braiding matrix $B$ proportional to the fusing matrix $F $
\eqnn{brai}
&& B_{\beta_1\!-\!h_s b,
\beta_2\!-\!h_t b}    
\left[\begin{matrix}\beta_1&\beta_2\cr  f &\b_3 \end{matrix} \right] (\epsilon)=
\\
&&
e^{i\pi  \epsilon( \tri(\b_3)\!+\!\tri(f)\!-\!\tri(\beta_1\!-\!h_s b)\!-\!\tri(\beta_2\!-\!h_t b))}
 F_{\beta_1\!-\!h_s b,
\beta_2\!-\!h_t b}
\left[\begin{matrix}\b_2& f\cr   \b_3&\b_1  \end{matrix} \right] \,. \nn
\eea
The generators  $e_1$ and $e_3\,,$ which exchange  the operators in the first two, respectively last two,  fixed points,    reduce 
 to  diagonal matrices. 

The   crossing symmetry  leads to a set of equations relating the 3-point constants and the fusing matrix elements
  \eqnn{loceq}
&&\sum_{h_s\in \G_{\o_2}}
 {c_{h_s}(\b_1)  C(\b_1 -h_s b, \b_2,\b_3)\over
 c_{h_t}(\b_2)\, C(\b_1, \b_2-h_t b, \b_3)}
 F_{\b_1 -h_s b, \b_2 -h_t b}\,
 F_{\b_1 -h_s b, \b_2 -h_u b} =\delta_{h_t, h_u}\cr
 &&= \sum_{h_s\in \G_{\o_2}} (F^{-1})_{h_t\, h_s} F_{h_s\, h_u}\,.
 \eea
 As discussed above we are left with summation over 3 of the  6 weights in the weight diagram
$ \G_{\o_2}$, as given in \eqref{weidi}.  
 A  shorthand notation for the matrix 
 $F_{h_s, h_t}= F_{\b_1 -h_s b, \b_2 -h_t b}$ in the last equality in \eqref{loceq} is used. 
The inverse matrix $F^{-1}$  is given by 
 \eqn{penta}
 (F^{-1})_{h_t , h_s}= F_{h_t, h_s}(\b_2,\b_1, \b_3)=  F_{\b_2 -h_t b, \b_1 -h_s b}
\left[\begin{matrix}\beta_1&-b\o_2\cr   \b_3  &\b_2  \end{matrix} \right] \,.
 \ee
According to \eqref{loceq} it can be identified with  the matrix formed by the ratio of constants times $F$, i.e., 
  \eqn{pent}
  {c_{h_s}(\b_1)  C(\b_1 -h_s b, \b_2,\b_3)\over
  c_{h_t}(\b_2)\, C(\b_2-h_t b, \b_1, \b_3)}
F_{h_s , h_t }(\b_1,\b_2, \b_3)= F_{h_t , h_s }(\b_2,\b_1, \b_3)\,.
 \ee

$\bullet$  
 Given the 3-point constants  \eqref{formT} and the OPE coefficients \eqref{weidi} one can solve the equations
  \eqref{penta},  \eqref{pent} for the $3 \times 3$ fusing matrix $F$.
We shall cast the solution of \cite{FP}   
    in a compact form  
     introducing first  some notation.  
  Define for weights of type $(\b_a, \a_1)=0=(\b_a, \a_3), a=1,2,3$  and  $\epsilon, \epsilon'= \pm 1$
\eqnn{Lfus}
&&G_{+,+}(\b_1,\b_2,\b_3)\equiv G_{+,+}(\b_1,\b_2,4Q \o_2-\b_3):=\\
&&{ \G(1+b(2\r Q-\b_1,\a_2))\, \G(b(\b_2-2\r Q, \a_2))\over
\G(1+b((Q+b) \o_2-\b_{13}^2, \o_1))\, \G(b(\b_{23}^1-\o_2 (Q+b), \o_1))\, }\,,  \nn \\
&& \nn\\
&&
G_{\epsilon,\epsilon'}(\b_1,\b_2,\b_3):=G_{+,+}(2(1-\epsilon) Q\o_2+\epsilon \b_1,2(1-\epsilon')Q \o_2+\epsilon '\b_2, \b_3)\,.  \nn 
\eea
Then with $s,s'=\pm 1$, referring to  the shifts $\b_1 \mp s \o_2 b$ resp. $\b_2 \mp s' \o_2 b$, 
and  $s, s' =0$,  referring   to $\b_1- \bh b$, resp.  $\b_2- \bh b$,   with $\bh$ defined in \eqref{weidi} we have 

\eqnn{pentagc} 
&&
F_{s,s'}\left[\begin{matrix}
\b_2&-b\o_2\\
\b_3&\b_1
 \end{matrix} \right] =         
\sum_{p=\pm, |p-2s'|=1}
{G_{-p,+}(\b_2,  {\o_2\over b}, \b_2+2s' Q\o_2)\over G_{u,+}(\b_1+2s Q\o_2,  {\o_2\over b}, \b_1)}\times \nn \\
&& \\
&&G_{-u,-p}(\b_1+(2s-u)Q\o_2, \b_2, \b_3)
G_{u-2s,p-2s'}(\b_1, \b_2+p Q\o_2, \b_3)\,.  \nn
\eea
Here 
$u=s$ if $s=\pm 1$, and   if $s=0$ one   can choose  one of the two values.

The  matrix  \eqref{pentagc} coincides   with the  $3 \times 3$ fusing matrix in the Liouville 
 theory
with one of the charges  given by the degenerate h.w. $\g=-\tilde{b}$, where  the parameter parameterizing the central charge $c_V=13+6(\tilde{b}^2+{1\over \tilde{b}^2}) $ is 
$\tilde{b}^2 =-Qb $ (i.e., $c_V < 1 $ for real $b$), namely:
\eqn{WfL}
F_{s,s'}\left[\begin{matrix}
\b_2&-b\o_2\\
\b_3&\b_1
 \end{matrix} \right]  =F^L_{\g_1+s{\tilde{b}},\g_2+s'{\tilde{b}} }\left[\begin{matrix} {\g}_2&-\tilde{b}\\ {\g}_3&{\g}_1  \end{matrix} \right] \,, \ s,s'=\pm 1, 0 \,, 
\ee
 under the identification of $W_2$ and $W_4$ highest weights 
\eqn{relfl}
2{\g}_a \tilde{b}= 1+Q b -(\b_a, \a_2)b= 2 (1 + (b\o_2-\b_a, \o_1)b)\,, \ a=1,2,3\,.
\ee
The reason for this coincidence will be cleared in the next Section.

\section{Integral representation for the 4-point function}
\setcounter{equation}{0}

The BF \cite{BF}  integral formula which allows to solve recursively some multiple Coulomb integrals representing 3-point functions
can be also exploited in order to give meaning of the so far formal 4-point correlator of Section  3. Namely 
one can interpret the latter as the analytic continuation of a 4-point Coulomb correlator  with the four charges 
constrained by a  charge neutrality condition 
- the latter is transformed,  achieving an alternative double integral representation of Coulomb gas type, 
so that the whole dependence on the finite  number of screening charges  
 is  located in the overall constant, which admits an analytic continuation in the standard way. The computation is analogous to the one for Liouville correlators in \cite{FLa}, but unlike that  case the type of the integral representation is not preserved. We obtain ($w=w_{121321}$)
\def\tg{\tilde{\gamma}}
\eqnn{equivrepr}  
&&\la V_{- \o_2 b } (x) V_{\b_1} (0)V_{\b_2}(1)V_{\b_3}(\infty)\ra =R_{w}(\b_3) 
\la V_{- \o_2 b } (x) V_{\b_1} (0)V_{\b_2}(1)V_{2\r Q-\b_3}(\infty)\ra \cr
&& \\
&&= 
\Omega(\{\b_a\})  
   |x|^{2b(4\r Q-\b_1,\a_2)}|x-1|^{2b(4\r Q-\b_2,\a_2)}\, 
   I_2(\tg_1, \tg_2,\tg_3)(x)\, \nn 
 \eea
 where $$I_2(\tg_1, \tg_2,\tg_3)(x)={1\over 2\pi^2} \int  \int  |t_1-t_2|^{-4\tilde{b}^2}
 \prod_{i=1}^2  |t_i|^{ -4\tilde{b} \tg_1 }
  |t_i-1|^{ -4\tilde{b} \tg_2 }|t_i-x|^{-4\tilde{b}\tg_3}d^2t_i
 $$ is the Liouville Coulomb integral with two screening charges,
 satisfying  
 \eqn{Lcons}{\(\tg_{1234}+2\tilde{b}\)\tilde{b} =  \tilde{Q}}\tilde{b}=-b^2\,. \ \ 
  \ee 
  They are related 
 to the $W_4$ weights $\b_a\,, a=1,2,3\,, \b_4=-b\o_2$ as
\eqnn{compLa}  
&&2\tg_{a}\tilde{b}=  -(2\b_a-\b_{124}^3, \o_1)b 
+2Qb \,, a=1,2\nn \\
&&2\tg_{3}\tilde{b}= (2\b_4-\b_{124}^3, \o_1)b
 +2Qb   \,.
\eea
The constant in the r.h.s. of \eqref{equivrepr}   is given by  
\eqnn{coflne}
&&\Omega(\{\b_a\}) =  { b^{4Qb-{2Q\over b}}\g(-{Q\over b})  
\over  \g(2Qb)  } \, 
{\(b^{2(1-b^2)}\pi \mu \g(b^2)\)^{ {(2\r Q-\b_{1234} ,\r)\over b}}\,
\zu^2(b)   \over  \zu((\b_{1234}, \o_1)-2Q) \zu((\b_{1234},\o_1)- 3Q )}\times
\nn \\
&& \nn  \\
&&\prod_{a=1,2,3} {\zu((\b_a,\a_2))  \zu((\b_a,\a_2)-Q) 
 \over  
\zu((\b_{1234} -2\b_a, \o_1)+b)  \zu((\b_{1234}-2\b_a,\o_1)-Q+b)} \,.
\eea
The integral in \eqref{equivrepr}  is equivalent \cite{FLa} up to a constant and an overall $x$ factor to the 
Liouville $4$-point function $\la V_{ - \tilde{b}}(x)  V_{ \g_1}(0) V_{\g_2}(1)  V_{\g_3}(\infty) \ra^L$ with three arbitrary weights $\g_a$  and one degenerate $-\tilde{b}$, in our notation 
\eqn{FLL}
\g_a=\tg_a+\tg_{3}+{\tilde{b}}\,, a=1,2\,, \ \g_3={1\over \tilde{b}}-\tg_1-\tg_2 \,.
\ee
The relation between the  highest weights of the vertex operators in the $W_4$  and the $W_2$ CFT  takes the universal  form 
\eqref{relfl}  if we furthermore change notation $\b_3 \to w_{2132}\bdot \b_3=4 Q \o_2 -\b_3$ in \eqref{equivrepr},  which leaves
invariant the expression  \eqref{relfl}  for the $F$ matrix elements. 

\section{Braiding identity}
\setcounter{equation}{0}
One checks that in all the products  
\eqn{FF}M_{h_s, h_t}(\b_1,\b_2, \b_3): = F_{h_s, h_t}(\b_1,\b_2,\b_3)F_{h_t, h_s}(\b_2,\b_1,\b_3)
\ee
the Gamma functions combine according to  $\G(z)\G(1-z)=\pi/\sin \pi z$,  so that \eqref{FF} turn into ratios of trigonometric functions. 
These values are actually solutions of an equation which 
arises from  the  braiding identity
\eqn{eqms}
\Omega_1 \Omega_2\Omega_3:= (e_1^2) (e_2 e_1^2 e_2^{-1}) (e_3 e_2 e_1^2 e_2^{-1}e_3^{-1}) (=
e_1 e_2 e_3^2 e_2 e_1)=  e^{-4\pi i \triangle(f)}\,.
\ee
The identity \eqref{eqms}  expresses the triviality up to phase of the composition of  monodromies  around 
the  three vertex coordinates: this is a property of the braid group on the sphere with 4 holes.  
The first and the last factors in the matrix relation  \eqref{eqms} are diagonal.
The eigenvalues $e^{2\pi i p(\b, h)}$ of the monodromy matrix  $\O$ are computed from the difference of Toda dimensions, cf.  \eqref{brai}
\eqn{eigvtoda}
p(\beta; h)= \triangle(\b - hb)-\triangle(\b)-\triangle(-b\o_2)= 2bQ+ b (\b-\r Q, h ) \,.
\ee
Multiplying   both sides of  \eqref{eqms} with $\O_3^{-1}$ and taking the trace one gets an equation for the products \eqref{FF}
 \eqn{janid}
{\sum_{h_s,h_t}}^{'}e^{2\pi i p(\b_1; h_s)}  M_{h_s, h_t}(\b_1,\b_2, \b_3)e^{2\pi i p(\b_2; h_t)} 
=e^{-4\pi i \triangle(f)}{\sum_{h}}^{'}e^{-2\pi i p(\b_3; h)} \,.
 \ee
In our case 
the summmation over the weight diagram $ \G_{\o_2}$ is restricted to the three points in \eqref{weidi}. The equation 
is checked to hold true for the explicit data presented in Section 3.

$\bullet$  In the limit $b\to 0$  the r.h.s of the matrix relation \eqref{eqms} becomes an identity  for any of the fundamental weights $f=-\o_i b$; in our case $4\tri(-\o_2 b) \to 0$ mod $2$. Taking this limit for 
  three "heavy" charges  $\beta_a = \eta_a/b$ with finite $\eta_a\,,  a=1,2,3$,   
one has   
 $p(\eta_a/b; h) \to (\eta_a,h)$ mod $2$. 
 
 The resulting  simplified identities  are precisely of the type  encountered  \cite{JW, KK}
in the  construction 
 of the quasiclassical 3-point functions of  sigma models (for the example of $AdS_3\times S^3$)
 in the framework of the AdS/CFT
 correspondence.
To make a connection with these considerations one identifies  the parameter $b^2$
 with the inverse of the 't Hooft coupling $b^2=1/\sqrt{\lambda}$.  
In the sigma model case the generators of the braid group 
 depend on the spectral parameter, i.e.,  $\eta=\eta(x)$,  and  $F=F(x)$. 
 However the 
$F(x)$  products  in  \eqref{FF}  solving the equation \eqref{janid}  in the quasiclassical limit  are identical as (trigonometric) functions of $\eta_a(x)$ to the ones found in the CFT.  Given these products,  
in the next step \cite{JW, KK} find  integral representations for the nontrivial individual $F(x)$ matrix elements with measure depending on the spectral curve.  We see that Toda  
 CFT (and more generally,   the WZW models, related to them  via the quantum Drinfeld-Sokolov reduction)  
 provide   important data on the string side of the constructions of the AdS/CFT
 correspondence.

\begin{footnotesize}
\section*{Acknowledgments}

VBP  would like to thank  for the  invitation the organizers 
of   the Workshop "Supersymmetries and Quantum Symmetries", 
Dubna,    2015. This  work has been partially supported by  the 
 Bulgarian NSF Grant   DFNI T02/6 and by the  
COST actions MP-1210 and MP-1405.

\end{footnotesize}

\end{document}